\documentclass[12pt]{iopart}
\usepackage{graphicx}

\begin{document}

\title[Electron pairing in the Hubbard model as a result of on-site repulsion fluctuations]{Electron pairing in the Hubbard model as a result of on-site repulsion fluctuations}

\author{Igor N.Karnaukhov}

\address{G.V. Kurdyumov Institute for Metal Physics, 36 Vernadsky Boulevard, 03142 Kiev, Ukraine}
\ead{karnaui@yahoo.com}
\vspace{10pt}
\begin{indented}
\item[]December 2020
\end{indented}

\begin{abstract}
We focus our quantitative analysis on the stability of insulator state in the Hubbard model at half-filling.
Taking into account macroscopic fluctuation of the on-site repulsion, we consider the possibility of realizing a steady state which is characterized by electron pairing.  Fluctuation of the on-site repulsion leads to the formation of holes in the form of excited states. The electron liquid has two possibilities of relaxation in the state with bare on-site repulsion U: trivial, to the initial state, and nontrivial to state in which electrons polarize holes  forming  electron-hole pairs. A steady state is determined by minimum energy for given U and its fluctuation $\delta U$. The values U and $\delta U$, for which the states with electron pairing are stable, are calculated. The proposed pairing mechanism is to some extent similar to the formation of a long-range pairing correlation in an optically induced Hubbard chain \cite{H1}.
\end{abstract}

%
%
%
%
%

\section{Introduction}

The Hubbard chain is unstable at half filling occupation, according to the Lieb-Wu solution \cite{LW} the fermion spectrum is gapped for arbitrary (non equal to zero) on-site repulsion \cite{LW,AA}. The authors noted absence of Mott transition in the 1D model \cite{LW}.  In 2D and 3D lattices at half filling, the Mott transition in insulator phase is realized at finite value of the on-site Coulomb repulsion \cite{K0}. The Hubbard model also draws attention to the possible realization of electron pairing due to the on-site Coulomb repulsion,  so-called $\eta$ pairing \cite{H1,Y}. Since the compounds, in which high-temperature superconductivity is realized,  have been discovered the question of non-trivial nature of superconductivity remains relevant.  Unfortunately, today we have quite exotic mechanisms of electron interaction to explain high-temperature superconductivity, which lead to electron pairing \cite{Y,1,1a,1b,2}.
A large number of new superconducting materials have been also discovered: high temperature cuprate superconductors, ruthenates, ferromagnetic superconductors,  organic materials. These materials clearly indicate that pairing occurs due to electronic correlations, unlike traditional superconductors.
Study of the stability of electron liquid state with respect to superconducting fluctuations (or electron pairing) can make it possible to propose a real pairing mechanism in high-temperature superconductors.

The gap in the Hubbard model is a result of hybridization of electrons from different bands (bands with opposite electron spin) with momenta \textbf{k} and \textbf{k}$+\overrightarrow{\pi}$ \cite{K0}. Due to the hybridization the number of electrons in each band is not conserved,  spontaneous symmetry is also broken \cite{K2,K1}.
The gap formation mechanism in the Hubbard model is similar to the $ \eta $ pairing proposed by Yang \cite {Y, Yur}. Unfortunately $\eta$ pairing is not realized in the framework of the Hubbard model with the on-site Coulomb repulsion. We believe that the nature of the gap in the Hubbard model can be more complied when fluctuations of the on-site Coulomb repulsion are taken into account. At half filling the low-energy excitations are holes, so the fluctuation of the on-site Coulomb repulsion leads to the formation of holes in the spectrum, which can form pairs with electrons (due to the polarization of holes by electrons). In this case, fluctuation of the on-site Coulomb repulsion is effective attractive interaction, which can form electron-hole pairs. Electron pairing is determined by both values of the bare on-site repulsion and its fluctuation. We use this idea to consider the stability of the Hubbard model in the framework of a mean field approach with electron pairing. As result, the electron spectrum is determined by two gaps, the values of the gaps in chain and square, cubic lattices are calculated.
We do not consider the nature of fluctuations of the on-site Coulomb repulsion. Using the exact diagonalization method for the Hubbard chain  \cite{H1}, the authors showed that optical pump can indeed lead to properties similar to superconductivity.

\section{Model}

We will analyze the behavior of fermions in the framework of the well-known Hubbard model, the Hamiltonian $ {\cal H}= {\cal H}_0+ {\cal H}_{int}$ is written  as
\begin{eqnarray}
&&
{\cal H}_0= -\sum_{<ij>}\sum_{\sigma=\uparrow,\downarrow}a^\dagger_{i,\sigma}a_{j,\sigma} - \mu \sum_{j}\sum_{\sigma=\uparrow,\downarrow} n_{j,\sigma},\nonumber \\
&&
{\cal H}_{int} = U \sum_{j} \left(n_{j \uparrow}-\frac{1}{2}\right)\left(n_{j,\downarrow}-\frac{1}{2}\right),
 \label{eq-H1}
\end{eqnarray}
where $a^\dagger_{j,\sigma} $ and $a_{j,\sigma}$ are the fermion operators on a site \emph{j} with spin $\sigma=\uparrow,\downarrow$, $n_{j,\sigma}=a^\dagger_{j,\sigma}a_{j,\sigma}$ denotes the density operator. The Hamiltonian (1) describes the hoppings of fermions between the nearest-neighbor lattice sites  with the magnitudes equal to unit, $\mu$ is the chemical potential. ${\cal H}_{int}$ term is defined by the on-site Coulomb repulsion with the value of $U$.

We shall analyze the phase state of the system at half filling occupation for arbitrary dimension $d=1,2,3$, which corresponds to $\mu=0$. We consider  macroscopic fluctuation of the on-site repulsion $ \delta {U}> 0 $ with a large (in comparison with the characteristic electron times) relaxation time. At a sufficiently large fluctuation, when $ \delta {U}> \Delta $ ($\Delta$ is a gap in the spectrum), the quasi-particle excitations in an upper Hubbard band are holes since at half-filling and  $\delta{U}=0$ a lower Hubbard band is full. Effective on-site interaction can lead to the formation of electron-hole pairs. The state of electron liquid relaxes to initial state with bare on-site repulsion U. The system can return to its initial state, determined by the Hamiltonian (1), or to another state (in which the pairing of electrons occurs), which also corresponds to the same on-site repulsion equal to ${U} $. Minimal energy of the system corresponds to realization of the most probable state. We consider the model Hamiltonian in the form (1) with the following Hamiltonian ${\cal H}_{int}$
\begin{eqnarray}
&&
{\cal H}_{int} = (U+\delta{U}) \sum_{j} \left(n_{j \uparrow}-\frac{1}{2}\right)\left(n_{j,\downarrow}-\frac{1}{2}\right)
 \nonumber \\  &&
-\delta{W} \sum_{j} \left(n_{j \uparrow}-\frac{1}{2}\right)\left(n_{j,\downarrow}-\frac{1}{2}\right),
\label{eq-H2}
\end{eqnarray}
where the first term with the constant $U+\delta U$ causes the formation of hole excitations in the electron spectrum, the second term with the constant $-\delta{W}$ leads to the formation of electron-hole pairs.

When formulating Hamiltonian (2), the following assumptions were made; first, the fluctuation of the local repulsion is uniform and has macroscopic dimensions; second, its relaxation time is longer than the characteristic electron times. The second assumption makes it possible to consider the problem using the adiabatic approximation. In fact, Hamiltonian (2) describes the behavior of a macroscopic cluster in a medium with Hamiltonian (1). In this case, the clusterization of the environment can occur gradually.
The nature of the Mott-Hubbard phase transition lies in understanding the formation of a gap in the Hubbard model at half filling. According to [4], this phase transition is similar to the Peierls transition. In the Hubbard model  the effective field has a phase $\pi$, the cell doubles, and the gap opens for critical values of the on-site repulsion.  The phase of the effective field is a new unknown parameter that corresponds to the minimum of action.  This approach allows one to study the Mott-Hubbard phase transition for an arbitrary dimension of the Hubbard model. This idea was used to solve Hamiltonian (2), where there are two effective fields with different phases.

In  (2) we have separated two processes: formation of holes in the electron spectrum and their polarization due to the attraction of electrons to them. We study instability of the Hubbard model induced by the formation electron-hole pairs at half filling occupation. It should be noted that the insulator phase disappears in this case since both Hubbard bands are partially filled.

\section{Ground state}
\begin{figure}[tp]
 \centering{\leavevmode}
     \begin{minipage}[h]{.475\linewidth}
\center{
\includegraphics[width=\linewidth]{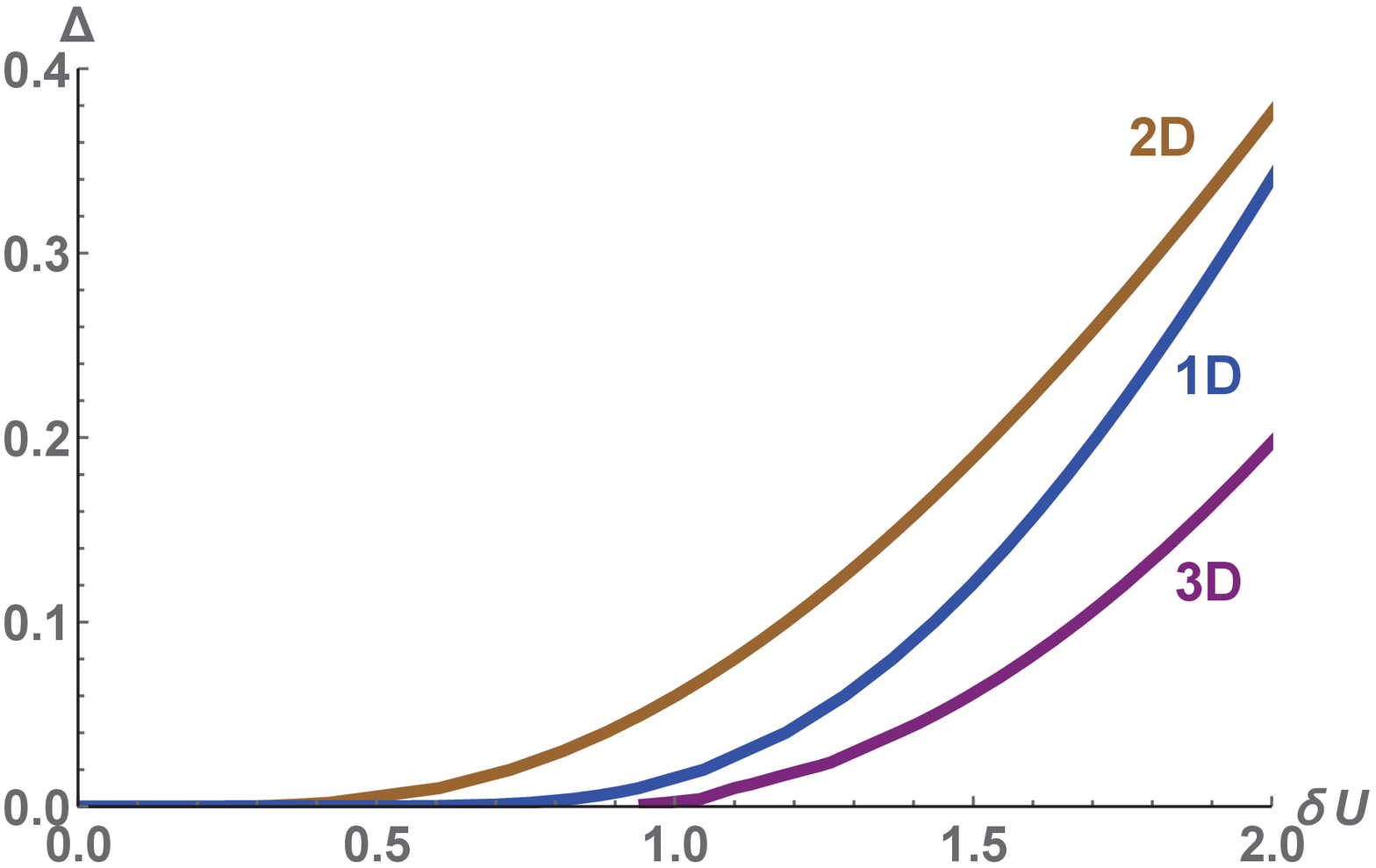} b)\\
                  }
    \end{minipage}
\centering{\leavevmode}
     \begin{minipage}[h]{.475\linewidth}
\center{
\includegraphics[width=\linewidth]{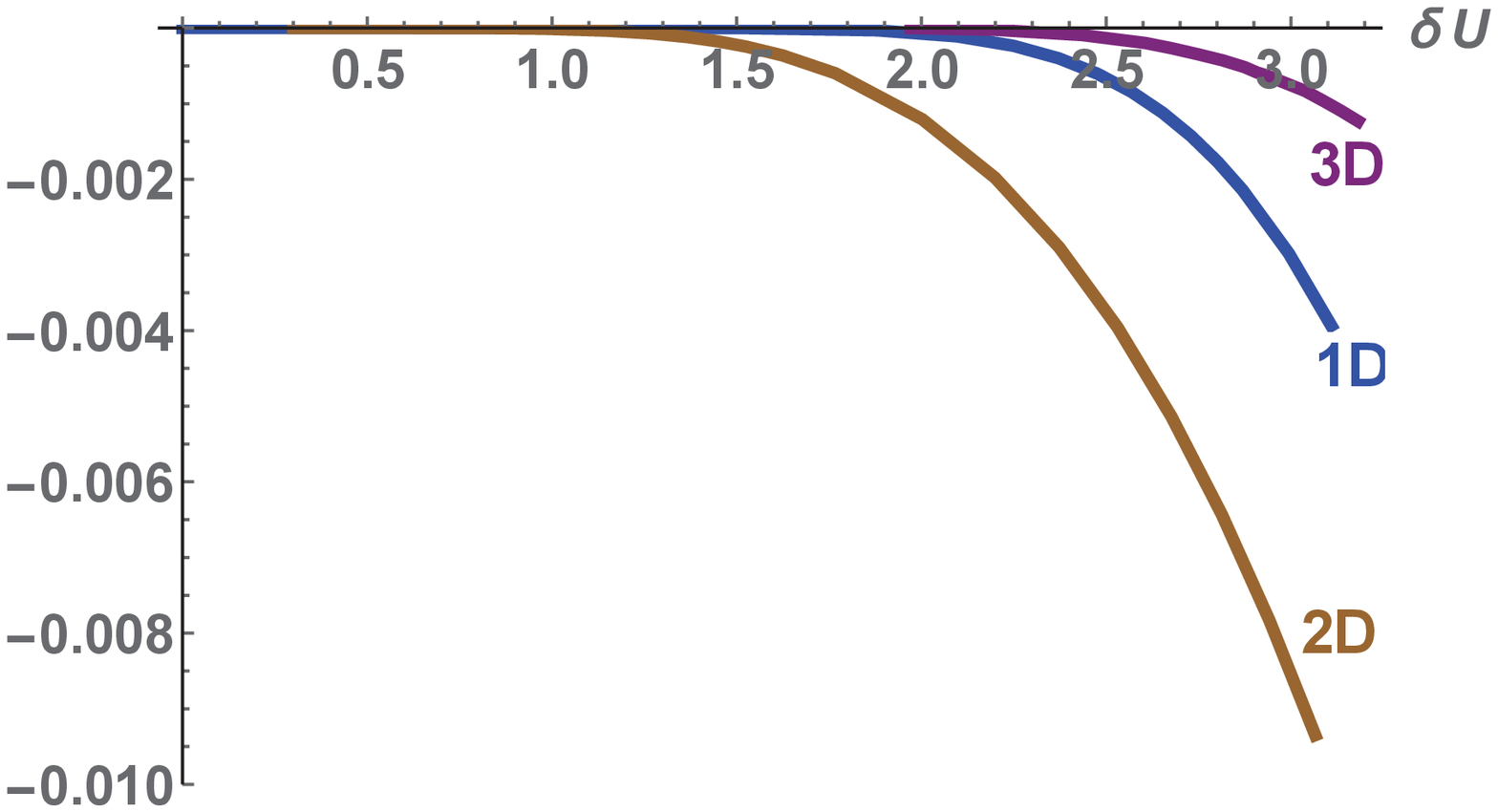} b)\\
                  }
    \end{minipage}
      \caption{(Color online) The gap $\Delta$ a) and the action $\delta S_{eff}$ b) as function of $\delta U$ at $U=0$ for different dimension d=1,2,3 (where $\lambda=\Lambda$ and $\Delta=\lambda+\Lambda)$.}
          \label{fig:1}
\end{figure}
\begin{figure}[tp]
  \centering{\leavevmode}
     \begin{minipage}[h]{.475\linewidth}
\center{
\includegraphics[width=\linewidth]{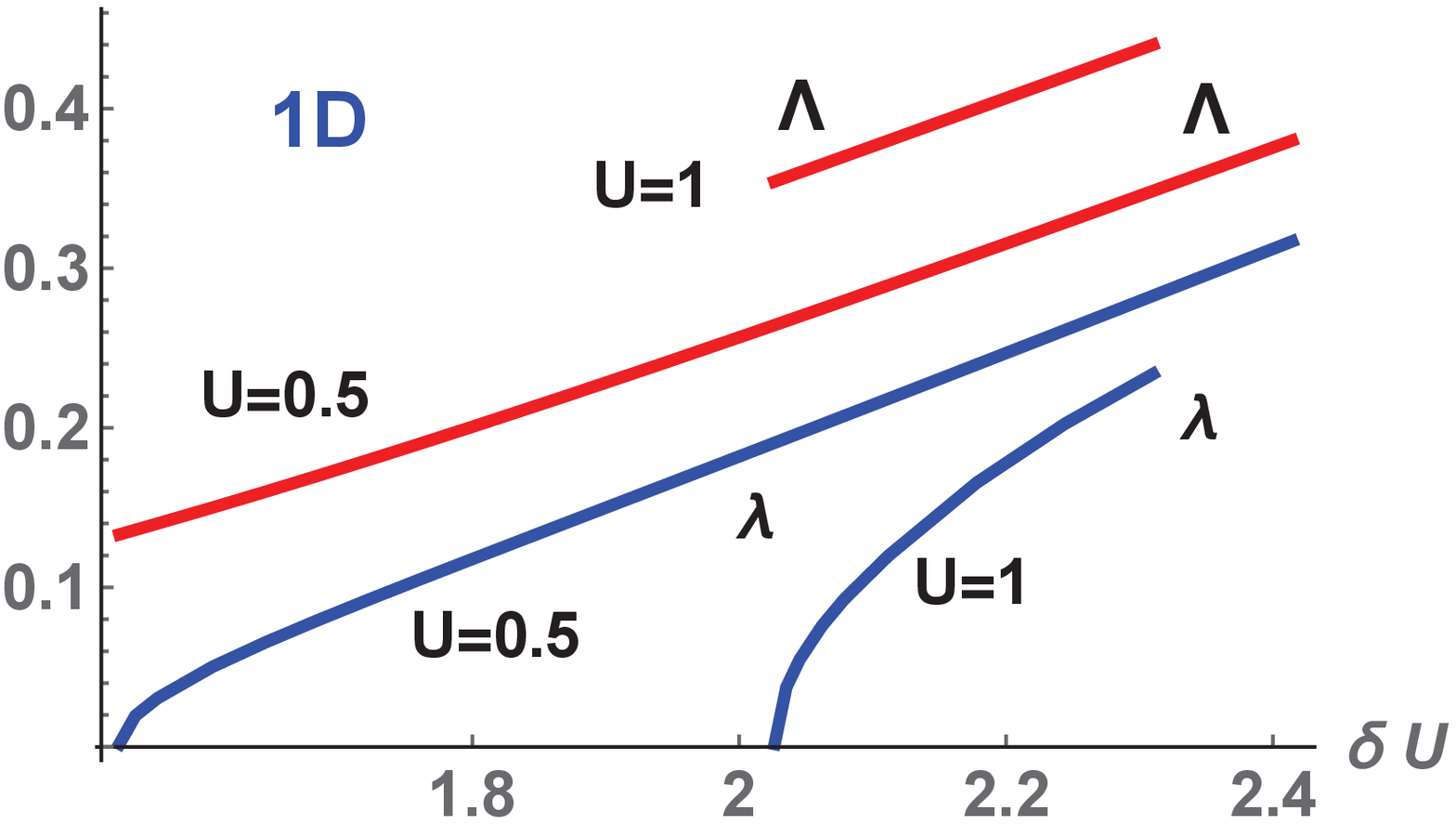} a)\\
                  }
    \end{minipage}
\centering{\leavevmode}
     \begin{minipage}[h]{.475\linewidth}
\center{
\includegraphics[width=\linewidth]{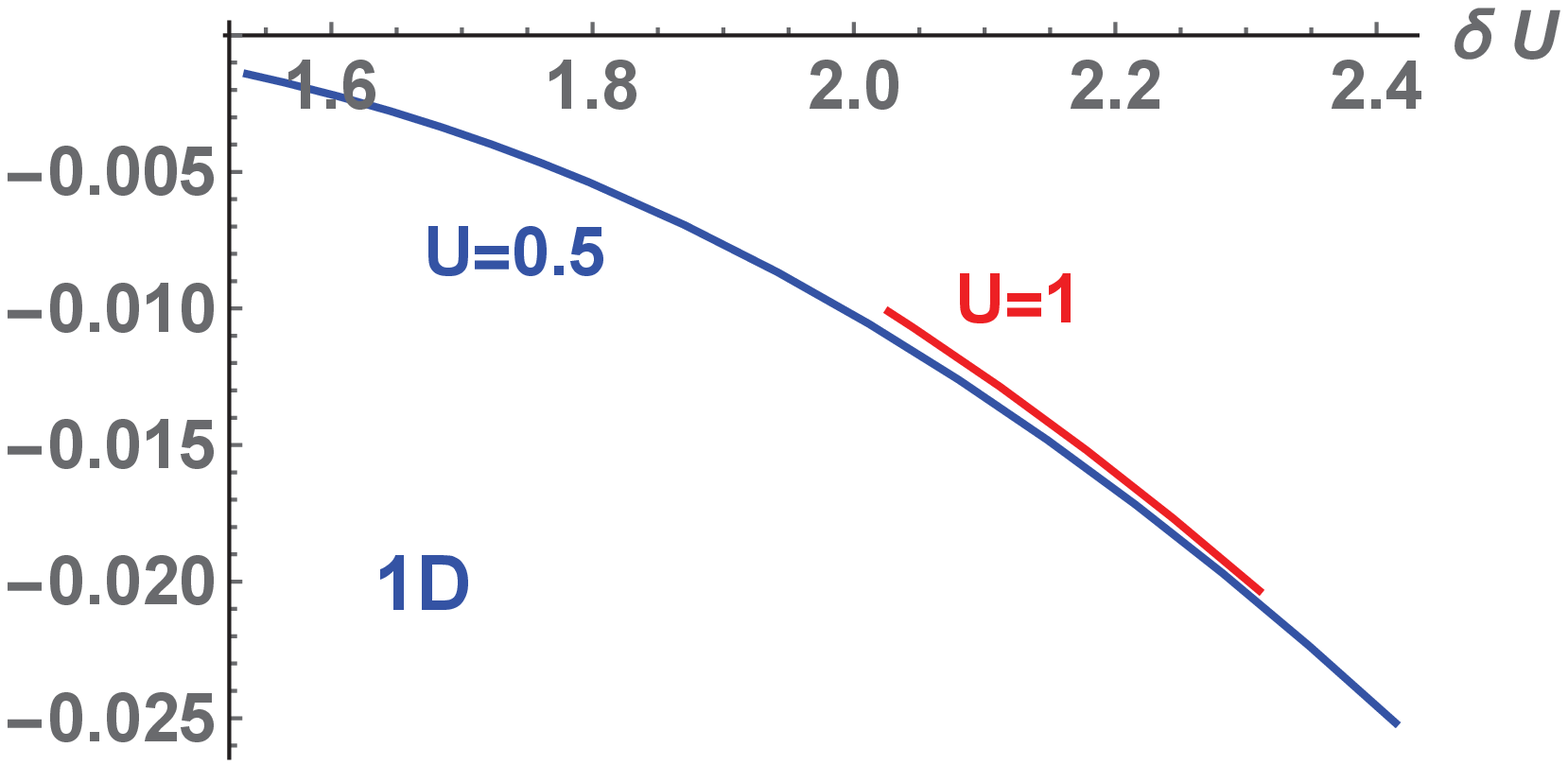} b)\\
                  }
    \end{minipage}
      \caption{(Color online) The components of $\lambda-\Lambda$-field a) and action $\delta S_{eff}$ b) as function of $\delta
      U$ calculated  at $U=0.5$ and $U=1$ for chain, where $\lambda=0, \Lambda=0.13$ at $\delta {U}=1.54$, when $U=0.5$ and $\lambda=0, \Lambda=0.354$ at $\delta {U}=2.02$ when $U=1$.}
          \label{fig:2}
\end{figure}

\begin{figure}[tp]
  \centering{\leavevmode}
     \begin{minipage}[h]{.475\linewidth}
\center{
\includegraphics[width=\linewidth]{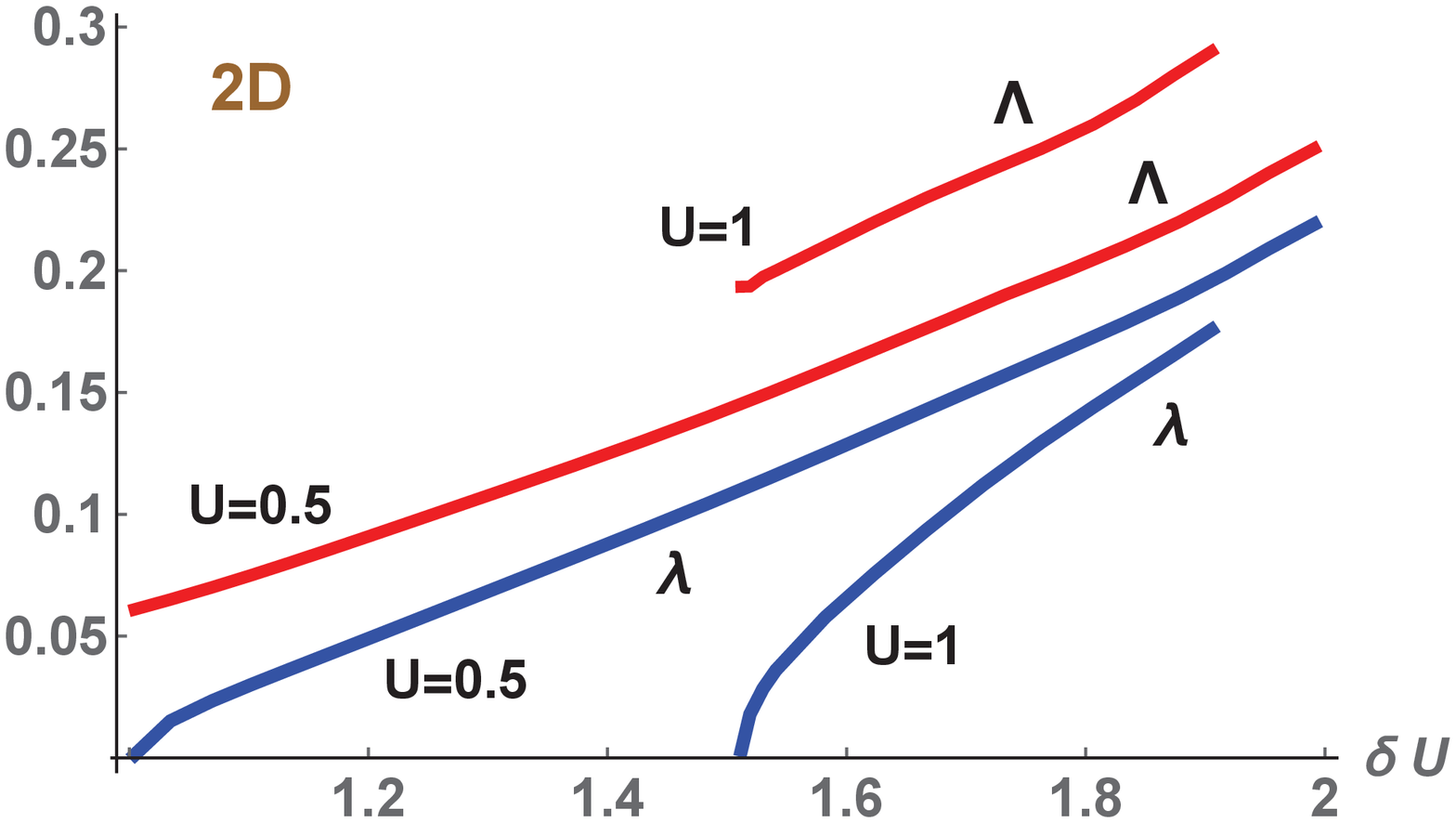} a)\\
                  }
    \end{minipage}
\centering{\leavevmode}
     \begin{minipage}[h]{.475\linewidth}
\center{
\includegraphics[width=\linewidth]{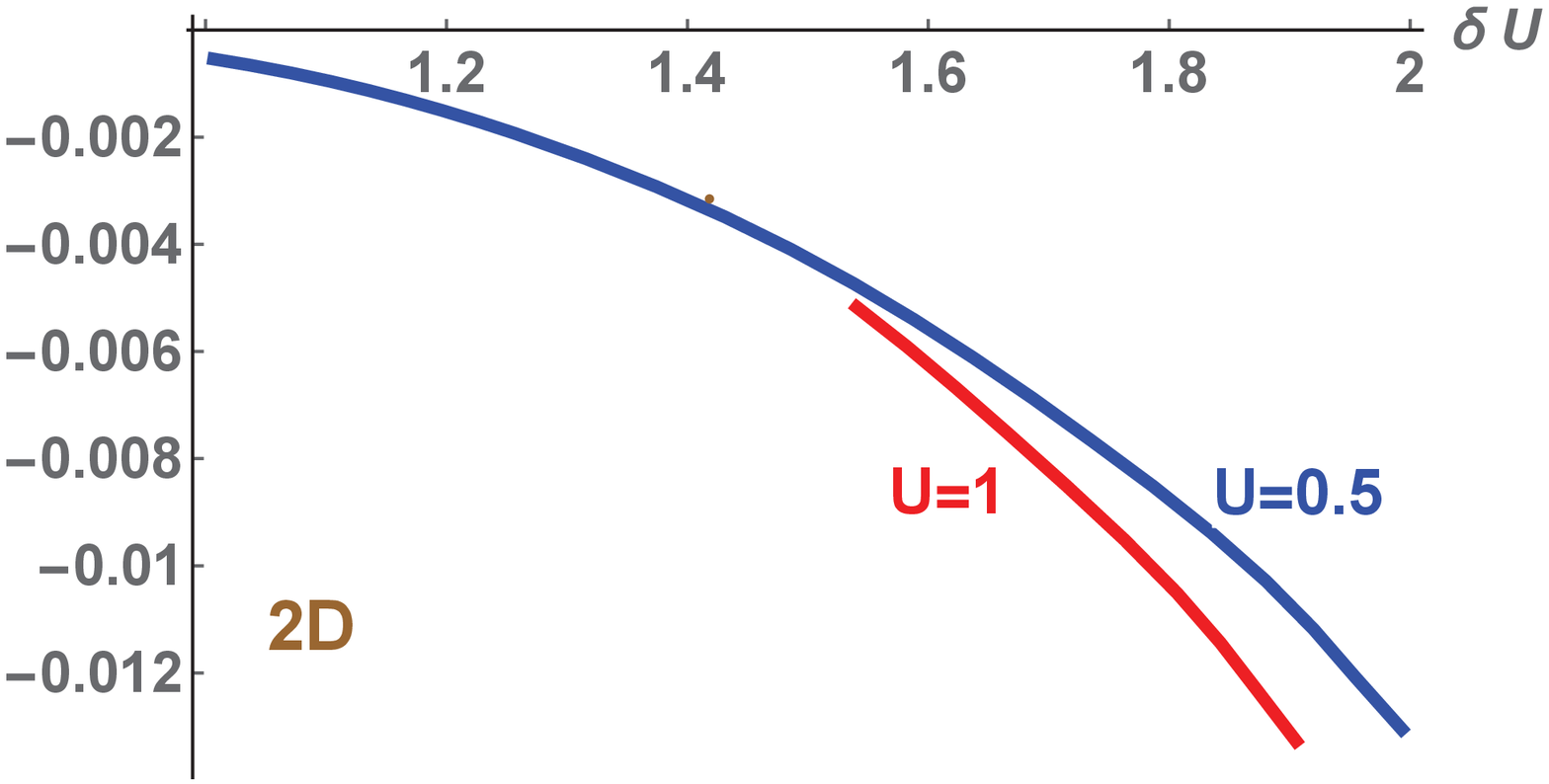} b)\\
                  }
    \end{minipage}
      \caption{(Color online) The components of $\lambda-\Lambda$-field a) and action $\delta S_{eff}$ b) as function of $\delta U$ calculated  at $U=0.5$ and $U=1$ for square lattice, where $\lambda=0, \Lambda=0.061$ at $\delta {U}=1$, when $U=0.5$ and $\lambda=0, \Lambda=0.193$ at $\delta{U}=1.51$ when $U=1$.}
          \label{fig:3}
\end{figure}
\begin{figure}[tp]
  \centering{\leavevmode}
     \begin{minipage}[h]{.475\linewidth}
\center{
\includegraphics[width=\linewidth]{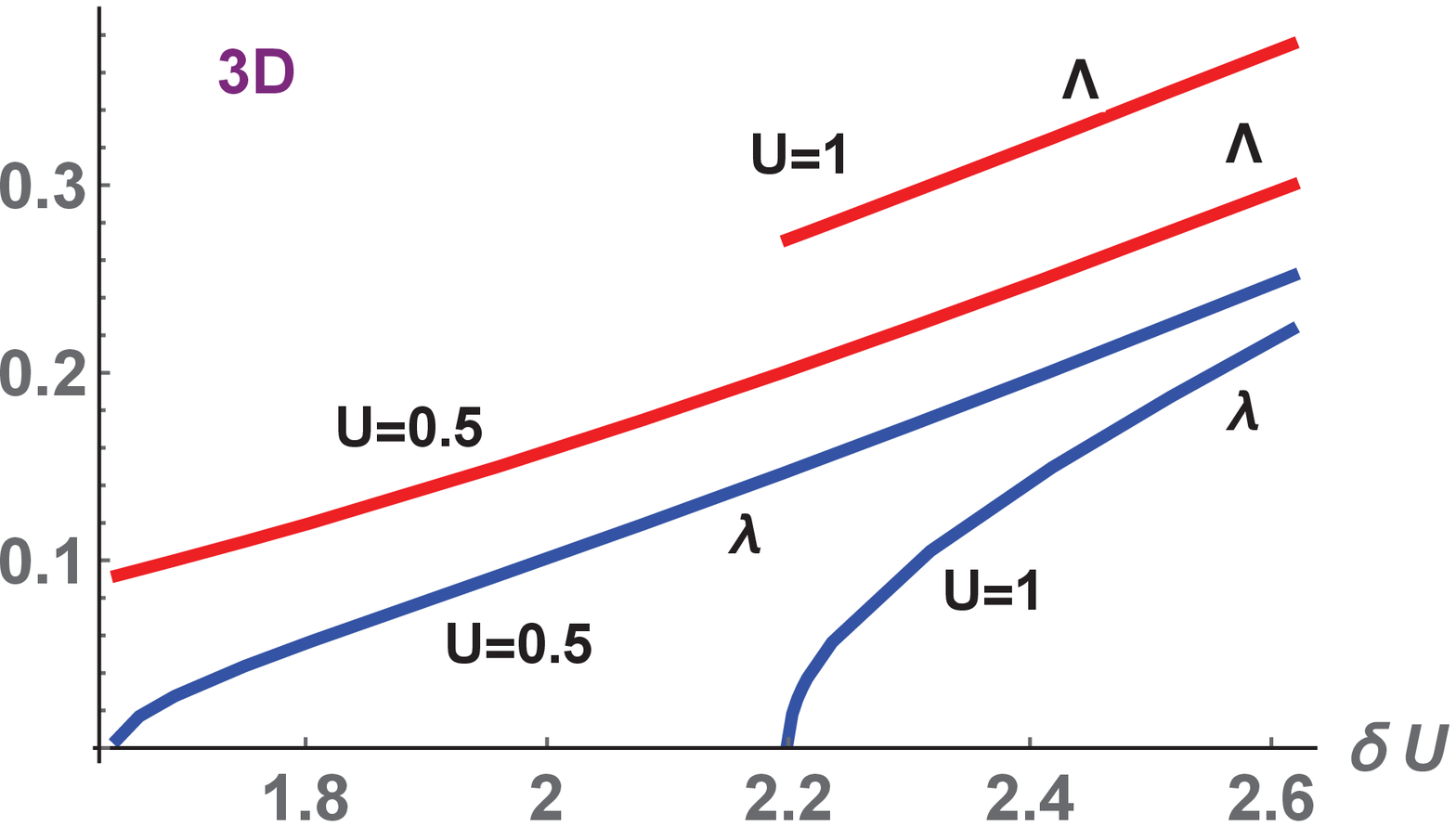} a)\\
                  }
    \end{minipage}
\centering{\leavevmode}
     \begin{minipage}[h]{.475\linewidth}
\center{
\includegraphics[width=\linewidth]{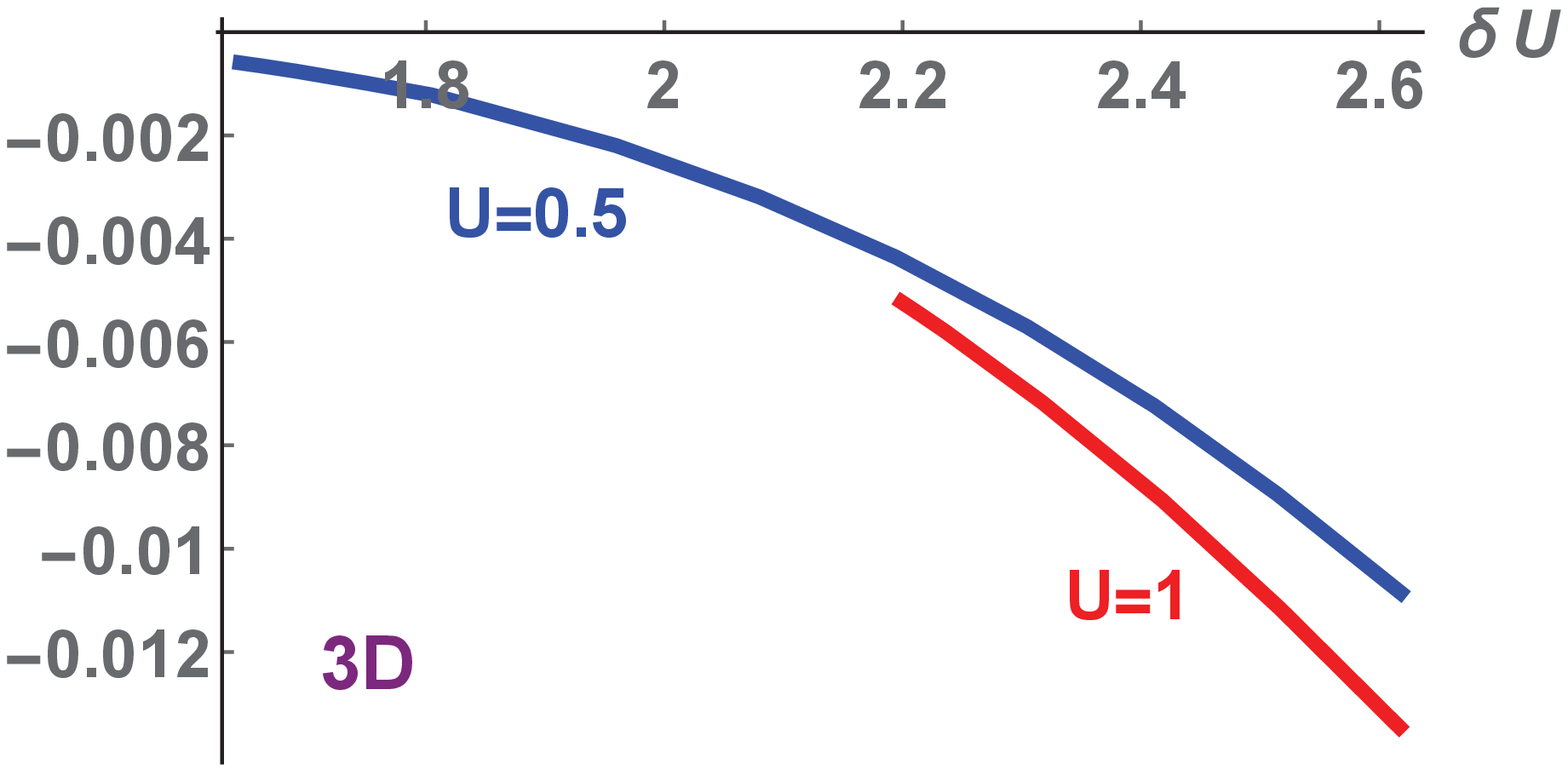} b)\\
                  }
    \end{minipage}
      \caption{(Color online) The components of $\lambda-\Lambda$-field a) and action $\delta S_{eff}$ b) as function of $\delta U$ calculated  at ${U}=0.5$ and ${U}=1$ for cubic lattice, where $\lambda=0, \Lambda=0.092$ at $\delta {U}=1.644$, when ${U}=0.5$ and $\lambda=0, \Lambda=0.271$ at $\delta {U}=2.2$ when ${U}=1$.}
          \label{fig:4}
\end{figure}
\begin{figure}[tp]
  \centering{\leavevmode}
     \begin{minipage}[h]{.475\linewidth}
\center{
\includegraphics[width=\linewidth]{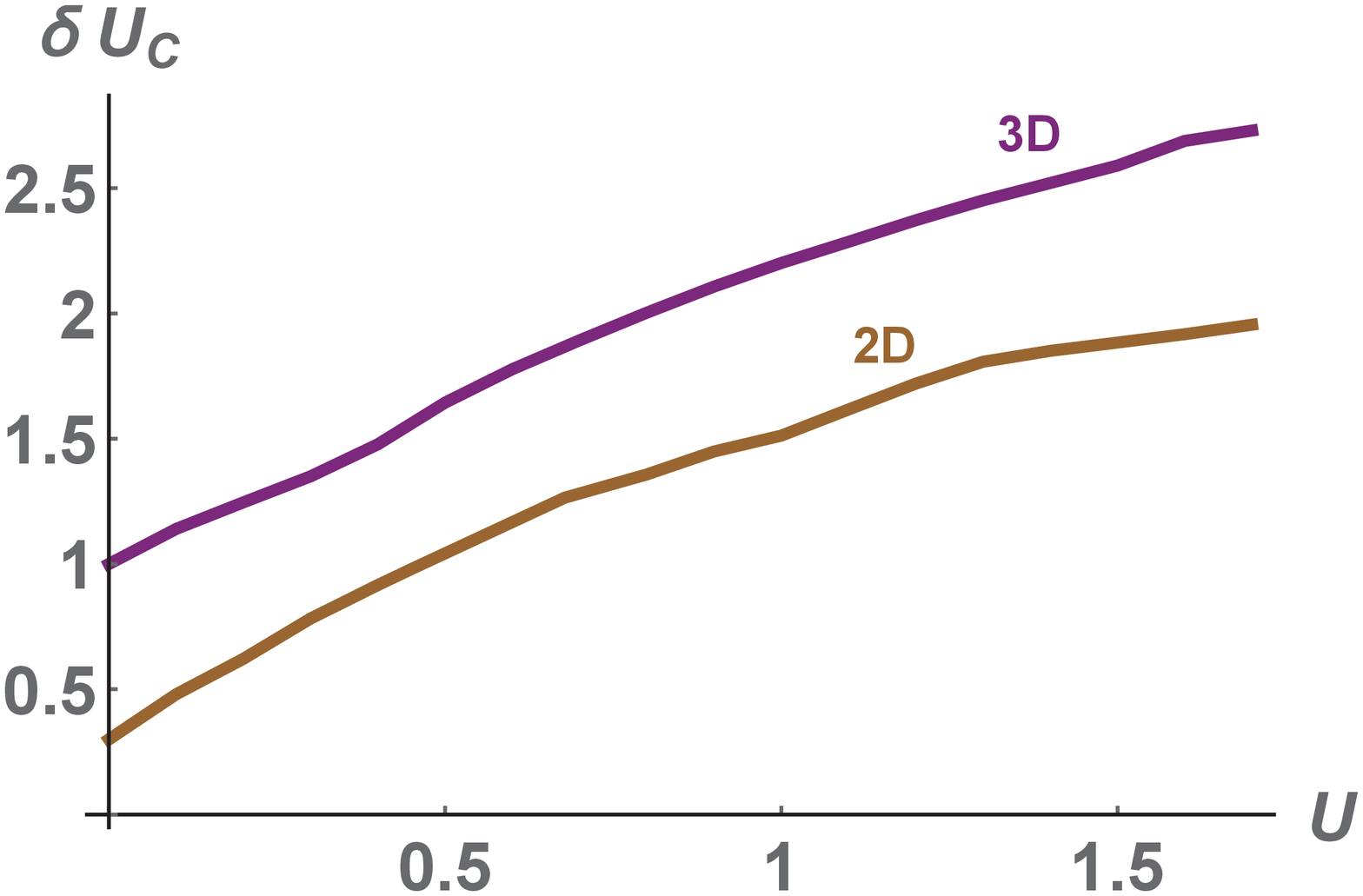}
                  }
    \end{minipage}
      \caption{(Color online) The minimal value of $\delta {U}$, above of which the state with nontrivial solutions for $\lambda$ and $\Lambda$ is realized, is calculated  as function of ${U}$ for square and  cubic lattices, where  $\delta {U}_c=0.3$ and $\delta {U}_c=1$ at ${U}=0$ in square and cubic lattices.}
          \label{fig:5}
\end{figure}

In section Appendix, we showed that in the mean field approximation, the behavior of electrons is described by their motion in the $ \lambda- \Lambda $ field. The solutions for $\lambda_j$ and $\Lambda_j$ are determined by unknown vectors $\textbf{q}$ and $\textbf{p}$, they determine the energies of the quasiparticle excitations $E_\alpha(\textbf{k},\textbf{q},\textbf{p})$ (see section Appendix). $\textbf{q}\neq 0$, $\textbf{p}\neq 0$ lift the degeneracy of the spectrum over the spin of electrons  forming a complex fermion spectrum. Due to symmetry of the fermion spectrum  the chemical potential is equal to zero at half-filling for arbitrary $\lambda, \textbf{q}$, $\Lambda, \textbf{p}$ and dimension of the model.
This makes it possible to compare the energies of the system for different parameters and fixed filling.
$\lambda-\Lambda$ field hybridizes the electron bands with different spins and momenta: $\lambda$-hybridization
with $\textbf{k}$ and $\textbf{k}+\textbf{q}$ momenta, $\Lambda$-hybridization
with $\textbf{k}$ and $\textbf{k}+\textbf{p}$ momenta. Thus, the electron-hole state with the momentum \textbf{k} is  tied to the states with momenta \textbf{k+q}, \textbf{k+p} and also \textbf{k+q+p} and \textbf{k+q-p}. Taking into account the spin freedom of electrons, the spectrum includes 16 branches. It allows us to calculate the state of the system which corresponds to minimum energy at $\mu=0$, the energy is equal to $E=\sum_{\alpha=1}^{16}\sum_{\textbf{k},E_\alpha(\textbf{k},\textbf{q},\textbf{p})<0} E_\alpha(\textbf{k},\textbf{q},\textbf{p})$. Numerical analysis  shows that the minimum of energy $E$ is always at the point  $\textbf{q}=\overrightarrow{\pi}$, $p=0$ (for arbitrary dimension of the model and $ \lambda,
\Lambda$). The ground state of the system  can be realized for $\textbf{q}=\overrightarrow{\pi}$, $p=0$, when pairs have zero momentum, with  following quasiparticle excitations
\begin{eqnarray}
&&
E_\pm^+({k})= \pm \sqrt{\frac{(\lambda+\Lambda)^2}{4} +\varepsilon_d^2(k)},
E_\pm^-({k})= \pm \sqrt{\frac{(\lambda-\Lambda)^2}{4} +\varepsilon_d^2(k)},
 \label{eq-0}
\end{eqnarray}

where $\varepsilon_d({k})=\sum_{i=1}^d \cos {k}_i$, $\textbf{k}=({k}_x,{k}_y,{k}_z)$. The spectrum is characterized by the gaps for each branch of excitations $\Delta_{\pm}=|\lambda \pm \Lambda|$.

The system relaxes to its initial state, which is described by the model Hamiltonian (1).
At $\delta W=\delta U$ the Hamiltonians (1) and (2) are determined by the same on-site Coulomb repulsion ${U}$, but in the second case this state can be formed by the excitations of electrons at $U+\delta{U}$ and pairing of electrons at $-\delta{U}$ due to fluctuation $\delta {U}$.

Given (3) in (7) we can obtain the following equations that correspond to the saddle point of action (7) at $\delta W=\delta U$
\begin{eqnarray}
&&
\frac{\lambda}{{U}+\delta{U}}-\frac{\lambda+\Lambda}{4{N}}\sum_{\textbf{k}}\frac{1}{\sqrt{(\lambda+\Lambda)^2 +4\varepsilon_d^2(k)}}-
\frac{\lambda-\Lambda}{4{N}}\sum_{\textbf{k}}\frac{1}{\sqrt{(\lambda-\Lambda)^2 + 4\varepsilon_d^2(k)}}=0,
\nonumber \\&&
\frac{\Lambda}{\delta{U}}-\frac{\Lambda+\lambda}{4{N}}\sum_{\textbf{k}}\frac{1}{\sqrt{(\Lambda+\lambda)^2 +4\varepsilon_d^2(k)}}-
\frac{\Lambda-\lambda}{4{N}}\sum_{\textbf{k}}\frac{1}{\sqrt{(\Lambda-\lambda)^2 +4 \varepsilon_d^2(k)}}=0.
\label{eq-1}
\end{eqnarray}

Action $S_{eff}(U,\delta U)$ (7) has the following form
\begin{eqnarray}
&&
\frac{S_{eff}(U,\delta{U})}{\beta}=\frac{\lambda^2}{{U}+\delta{U}}+
\frac{\Lambda^2}{\delta {U}}-
\frac{1}{2{N}} \sum_{\textbf{k}}(\sqrt{(\lambda+\Lambda)^2 +
4\varepsilon_d^2(k)}+\nonumber\\&&
\sqrt{(\lambda-\Lambda)^2 +4\varepsilon_d^2(k)}),
\label{eq-2}
\end{eqnarray}
where unknown $\lambda$ and $\Lambda$ are solutions of Eqs  (4).

\subsection{The case U=0}

$ \lambda $ and $ \Lambda$ define different physical processes in the system. $ \lambda $ defines the gap in the Hubbard model (for $ \Lambda = 0 $, this gap is equal to $ \lambda $ \cite {K0}), and $\Lambda $ defines electron pairing or polarization of holes by electrons. The magnitude of  $\delta U$, at which electron pairing occurs, cannot be less than the gap in the spectrum, because only in this case the  fluctuation forms holes. This also follows from the results of calculations, therefore, in 2D and 3D systems, the value of $\delta U $ cannot be infinitesimal.

Eqs (4) have both trivial solution $\lambda=\Lambda=0$, which correspond to noninteracting electrons, and nontrivial solution $\lambda=\Lambda\neq 0$, the solution that is determined by the effective repulsion $\delta U$ and attraction $-\delta U$ between electrons with different spins. Nontrivial solution for $\lambda\neq 0$ and $\Lambda\neq 0$  splits the branches of the spectrum of the Hamiltonian ${\cal H}_0$ (degenerated by the spin of noninteracting electrons) on gapless $E_\pm^-({k})=\varepsilon_d({k})$ and gapped  $E_\pm^+({k})= \pm\sqrt{\lambda^2+ \varepsilon_d^2 ({k})}$ with the gap $\Delta=2\Lambda=2\lambda$ (3) (see in Fig 1a ).

In the chain the gap state is realized for an arbitrary value of fluctuation $\delta U$. In weak $\delta U$ limit the gap is exponential small $\Delta = 4 G \exp(-2\pi/\delta U)$, where $G$ is cutoff that defines the region of integrating  for the momentum near $\mu = 0$. In square and cubic lattices nontrivial solution for $\lambda$, $\Lambda$ takes place at finite $\delta U >\delta U_c$, where $\delta U_c$ is a minimal value of fluctuation of on-site Coulomb repulsive at which the electron-hole states are realized. The value of $ \delta U_c$ determines the phase stability criterion.  This fact leads from numerical calculation, such  $\delta U_c=0.3$ for square and $\delta U_c=1$ for cubic lattices. $\delta U_c=0$ is not surprising in the chain because the same behavior of the fermion spectrum takes place in the Hubbard chain \cite{LW}.

The phase transition is accompanied by a decrease in action, which leads to the formation of a stable  new phase. Numerical calculations $\delta S_{eff}=S_{eff}(U,\delta U)-S_{eff}(U,0)$ as function of $\delta U$ at $U=0$ for different dimension of the system are shown in Fig 1b. For $ U = 0 $, the solution $ \lambda = \Lambda $ follows from equations (1). In this case, the action is determined by only one parameter, as a result, a non-trivial solution for $ \lambda $ and $ \Lambda$ corresponds to a global minimum of the action.

\subsection{The cases U=0.5 and U=1}

In this section, we consider the formation of the ground state of the model for bare Coulomb repulsion $U = 0.5 $ and $ U = 1 $ and an arbitrary dimension of the system. Nontrivial solutions for $\lambda$ and $\Lambda$ determine the ground state of the interacting electrons, these solutions take place at finite values of $\delta U$. A minimal value of $\delta U$, at which the state with electron pairing is realized, increases with value of a bare repulsion $U$. We illustrate these calculations at $U=0.5$ and $U=1$ for different dimension of the system (see in Figs 2,3,4). Numerical calculations of the averages $\lambda$ and $\Lambda$ and action $\delta S_{eff}$ as function of $\delta U$ are shown in Figs 2,3,4. The behavior of the system is similar for different dimension, the main result is as follows: the state, in which the electron pairing is realized, is stable and can be realized. In Fig 5 we have calculated the ground state phase diagram in the coordinates $U$, $\delta U$ for square and cubic lattices. The curves separate the regions in which electron pairing  occurs (above the curves) and not (below the curves).

\section{Conclusions}
The discovered instability of the Hubbard model at half filling occupation allows us to propose a possible mechanism of electron pairing. To realize such a pairing mechanism, sufficiently large macroscopic fluctuations of the on-site Coulomb repulsion are required, which can occur in low-dimensional systems. At T = 0, fluctuations in the superconducting order parameter kill the superconducting state in the chain. We have shown that any fluctuations in the on-site Coulomb repulsion lead to pairing of electrons in the chain of noninteracting fermions. Thus, there are two mutually exclusive  fluctuation processes that affect the formation (destruction) of electron pairing in the chain of noninteracting fermions.
Numerical analysis shows that fluctuations of on-site Coulomb repulsion $\delta U$ should be of the order of the magnitude of the bare Coulomb repulsion $U$, in the strong repulsion limit $\delta U \rightarrow U$. A chain of noninteracting electrons is unstable with respect to fluctuations of the on-site Coulomb repulsion, since any fluctuation in magnitude opens a gap at half-filling, leads to the formation of electron-hole pairs. Many high-temperature superconductors  have an effective dimension of two, which is preferred in this case. The gap in the electron spectrum is determined by the magnitude of the fluctuation of the Coulomb repulsion, therefore, we are talking about a large value of the gap and the electron-electron mechanism of pairing.

\section{Acknowledgments}

The studies were also supported by the National Academy of Sciences of Ukraine within the budget program 6541239 "Support for the development of priority areas of scientific research".

\section{Appendix}
Let us introduce the operators $\chi_{j}^\dagger= a^\dagger_{j, \uparrow}a_{j,\downarrow}$ and $\eta_{j}^\dagger= a^\dagger_{j,\uparrow}a^\dagger_{j,\downarrow}$ and redefine the term ${\cal H}_{int}$ (2) is the following form
${\cal H}_{int}= -({U}+\delta {U})\sum_{j}\chi_{j}^\dagger\chi_{j}-\delta{W}\sum_{j}\eta^\dagger_{j}\eta_{j}$. The Hubbard-Stratonovich transformation maps interacting fermion systems to non-interacting fermions moving in an effective field, we define the interaction term introducing the action ${S}_{0}$
\begin{eqnarray}
&& S_{int}=S_{0}+\sum_{j}\left( \frac{\lambda^\ast_{j}\lambda_{j}}{{U}+\delta{U}}+ \frac{\Lambda^\ast_{j}\Lambda_{j}}{\delta{W}}\right)+
\sum_{j}(\lambda_{j} \chi_{j}+\Lambda^\ast_{j}\eta^\dagger_{j}+ H.c.)
 \label{A1}
\end{eqnarray}

The canonical functional is defined as $${\cal Z}=\int {\cal D}[\lambda, \Lambda] \int {\cal D}[\chi\dagger,\chi,\eta^\dagger,\eta] e^{-S},$$ where the action $S=\frac{1}{{U}+\delta{U}}\sum_{j}\lambda^\ast_{j}\lambda_{j}+ \frac{1}{\delta{W}}\sum_{j}\Lambda^\ast_{j}\Lambda_{j}+
\int_0^\beta d\tau \Psi^\dagger (\tau)[\partial_\tau  + {\cal H}_{eff}]\Psi (\tau)$ with
$${\cal H}_{eff}=  {\cal H}_0 +\sum_{j}(\lambda_{j}a^\dagger_{j,\downarrow}a_{j,\uparrow}+
\Lambda^\ast_{j}a^\dagger_{j,\uparrow} a^\dagger_{j,\downarrow}+{H.c.}),$$ where $\Psi (\tau)$ is  the wave function. We expect that $\lambda_{j}$ and $\Lambda_{j}$ are independent of $\tau$ because of translational invariance.

At the on-site hybridization and on-site pairing, and due to translation invariance, only the phases of $\lambda_j$ and $\Lambda_j$ are depend on $j$, a namely $\lambda_{j}=\exp(i \textbf{q }\textbf{j}) \lambda$ and $\Lambda_{j}=\exp(i \textbf{p}\textbf{j}) \Lambda$, where $\textbf{q}$ and $\textbf{p}$ are unknown wave vectors. The task is reduced to moving fermions in a static inhomogeneous $\lambda-\Lambda$ field.
We can  integrate out fermions to obtain  the following action $S_{eff}$ per an atom (6) (N is the total number of atoms, a lattice constant is equal to 1)
\begin{eqnarray}
&&
\frac{S_{eff}(U,\delta U)}{\beta}=-\frac{{T}}{{N}}\sum_{\textbf{k}}\sum_n \sum_{\alpha=1}^{16} \ln [-i \omega_n+E_\alpha(\textbf{k},\textbf{q},\textbf{p})]
+\frac{|\lambda|^2}{{U}+\delta{U}}+
\frac{|\Lambda|^2}{\delta{W}},
 \label{A2}
\end{eqnarray}
where $\omega_n =T(2n+1)\pi$ are Matsubara frequencies, $\textbf{k}$, $\textbf{q}$ are the momenta of electrons, $\textbf{p}$ is the momentum of Cooper pair, 16-quasiparticle excitations $E_\alpha(\textbf{k},\textbf{q},\textbf{p})$ ($\alpha =1,...,16$) determine the electron states in the  $\lambda-\Lambda$ field.
In the saddle point approximation the canonical functional ${\cal Z}$ will be dominated by the minimal action $S_{eff}$ (7), that satisfies the following conditions $\partial S_{eff}/\partial \lambda =0$ and $\partial S_{eff}/\partial \Lambda =0$.

\subsection*{Author contributions statement}
I.K. is an author of the manuscript
\subsection*{Additional information}
The author declares no competing financial interests.
\end{document}